\documentstyle[12pt]{article}
\begin{document}
\title{INCONSISTENCY OF CANONICALLY QUANTIZED N=1 SUPERGRAVITY?
\thanks{Alberta Thy-28-93,
hep-th/9306032}}
\author{ Don N. Page\\CIAR Cosmology Program\\Theoretical Physics
Institute\\
Department of Physics\\University of Alberta\\Edmonton,
Alberta\\Canada T6G
2J1\\
Internet:  don@page.phys.ualberta.ca}
\date{1993 June 4}
\maketitle
\large
\begin{abstract}
\baselineskip 20pt
D'Eath's proof that there can be at most two allowed quantum states
of
$N=1$ supergravity with zero or a finite number of fermions can be
extended to show that there are no such states.
\end{abstract}
\normalsize
\pagebreak
\baselineskip 22pt

The canonical  (Dirac) quantization of $N=1$ supergravity \cite{D84}
applies
Lorentz and supersymmetry constraints
	\begin{equation}
	J^{AB}\Psi=\bar{J}^{A'B'}\Psi=S^A\Psi=\bar{S}^{A'}\Psi=0
	\end{equation}
to a wavefunctional $\Psi$ of the spatial tetrad $e^{A
A'}_{~~~~i}(x)$
and of one-half of the spatial gravitino field, say $\psi^A_{~~i}
(x)$.
D'Eath \cite{D93} argued that since the supersymmetry
constraints do not mix fermion number, one can study separately the
states
with definite fermion numbers.  Furthermore, for a nonzero finite
number
of fermions, he noted that the two supersymmetry constraints are
inconsistent,
so that only bosonic states are allowed.  Finally, D'Eath showed that
only
two bosonic states are possibly allowed, of the form $e^{\pm I}$ for
a certain
action functional $I$ of the three-geometry.

Bosonic wavefunctionals either have the form $ \Psi ( e^{A
A'}_{~~~~i} (x),~\psi^A_{~~i} (x) ) $ with no dependence on the
$\psi^A_{~~i} (x)$ half of the spatial gravitino field, or they have
the
fermionic Fourier transform \cite{D84}
 $ \tilde \Psi ( e^{A A'}_{~~~~i} (x),~\tilde \psi^{A'}_{~~i}(x) ) $
 with no dependence on the other half, $\tilde \psi^{A'}_{~~i}(x)$,
 of the spatial gravitino field.  In the former case, the
supersymmetry
 constraint can be written as \cite{D93}
 	\begin{equation}
	\Psi^{-1}\bar{S}_{A'}\Psi\equiv\epsilon^{i j k} e_{A A' i}
(\;^{3 s} D_j
	\psi^A_{~~k} ) - \frac{1}{2} \hbar \kappa^2
	\psi^A_{~~i} {\delta (\ln \Psi ) \over \delta e^{A
A'}_{~~~~i}} = 0~.
	\end{equation}

D'Eath derived a formula, Eq. (24) of \cite{D93}, for the variation
of
$\ln \Psi$ produced by a variation of the spatial tetrad
$e^{A A'}_{~~~~i}(x)$.
However, one can readily see that although this may make
$\delta (\ln \Psi ) / \delta e^{A A'}_{~~~~i}$ satisfy Eq. (2) for
{\it some}
possible spatial gravitino field configurations $\psi^A_{~~i} (x)$,
it cannot possibly satisfy Eq. (2) for {\it all} $\psi^A_{~~i} (x)$.
In particular, fixing a bosonic wavefunctional $ \Psi (
e^{AA'}_{~~~~i} (x) )$,
fixing the tetrad field $e^{AA'}_{~~~~i} (x)$, fixing the variation
$\delta e^{A A'}_{~~~~i}$, and fixing $\psi^A_{~~i} (x)$ at the one
point $x$
fixes the second term of $\Psi^{-1}\bar{S}_{A'}\Psi$ in Eq. (2) at
that point.
However, this still allows freedom to choose the 3-dimensional
covariant
derivative $\;^{3 s} D_j\psi^A_{~~k}$ of the arbitrary fermionic
field
$\psi^A_{~~k} (x)$ at the point $x$, so that generically the first
term of
$\Psi^{-1}\bar{S}_{A'}\Psi$ would not cancel the second term.
Therefore, Eq. (2) has no bosonic solutions ($\Psi$ independent of
$\psi^A_{~~k}$) for all $\psi^A_{~~k}$.

A precisely analogous argument applies to $S_A\Psi=0$ for the other
type of bosonic wavefunctional,
$ \tilde \Psi ( e^{A A'}_{~~~~i} (x),~\tilde \psi^{A'}_{~~i}(x) ) $
independent of $\tilde \psi^{A'}_{~~i}(x)$.  Thus it appears that
both
types of bosonic wavefunctionals are inconsistent with the
supersymmetry
constraints.  When this result is combined with D'Eath's argument
\cite{D93} against states with a finite nonzero number of fermions,
we see that all quantum wavefunctionals of $N=1$ supergravity
with a finite (zero or nonzero) number of fermions are inconsistent
with the supersymmetry constraints.

It is tempting to conclude from these results that cononically
quantized
$N=1$ supergravity is inconsistent.  However, D'Eath warns
\cite{D93},
``The case with an infinite number of fermions might of course be
different."
It would be interesting to see whether the arguments can be extended
to
that case.  Naively one would not expect a time-dependent geometry to
produce purely a bounded number of fermions, since there should be an
amplitude to produce an arbitrarily large number.  On the other hand,
at
least if the theory were finite without renormalization, one would
also
expect finite amplitudes for components of the wavefunctional with
finite
numbers of fermions.  By D'Eath's arguments and the additional
argument above, these components would be decoupled from each
other, and each would
apparently be inconsistent, unless somehow their inconsistency were
canceled by terms from the components with an arbitrarily large or
infinite
number of fermions.  Or, it might turn out that the physical states
are
entirely concentrated on components with an infinite number of
fermions.
In that case it would appear that one would need a more sophisticated
mathematical formulation of the canonically quantized theory.

One might also consider the possibility that $N=1$ supergravity
should not
be canonically quantized in the standard way \cite{D84}, or that
perhaps
it can be quantized only in a noncanonical way.  $N=1$ supergravity
might
be a theory that cannot be consistently quantized, or it might be a
theory
whose consistent quantization teaches us interesting lessons about
canonical or other methods of quantization.  In either case, we may
have
much to learn from this theory.

{\bf Acknowledgments}:  The hospitality of William and Barbara Saxton
at their home in State College, Pennsylvania, where the first draft
of this
paper was written, is gratefully acknowledged.  Discussions and
correspondence with Peter D'Eath (starting at the Journ\'{e}es
Relativistes
'93 at the Universit\'{e} Libre de Bruxelles, and with one fax kindly
sent
for me by Penn State University) have been very helpful.  Financial
support was provided in part by the Natural Sciences and Engineering
Research Council of Canada.

\end{document}